\documentclass[aps,prl,twocolumn,superscriptaddress,showpacs]{revtex4}
\usepackage{amsmath}
\usepackage{epsfig}

\input{tcilatex}

\begin{document}

\title{Non-Markovian band-edge effect and entanglement generation of quantum
dot excitons coupled to nanowire surface plasmons }
\author{Guang-Yin Chen}
\affiliation{Institute of Physics, National Chiao Tung University, Hsinchu 300, Taiwan}
\author{Yueh-Nan Chen}
\email{yuehnan@mail.ncku.edu.tw}
\affiliation{Department of Physics and National Center for Theoretical Sciences, National
Cheng-Kung University, Tainan 701, Taiwan}
\author{Der-San Chuu}
\affiliation{Department of Electrophysics, National Chiao-Tung University, Hsinchu 300,
Taiwan}
\date{\today}

\begin{abstract}
The radiative decay of quantum dot (QD) excitons into surface plasmons in a
cylindrical nanowire is investigated theoretically. Maxwell's equations with
appropriate boundary conditions are solved numerically to obtain the
dispersion relations of surface plasmons. The radiative decay rate of QD
excitons is found to be greatly enhanced at certain values of the exciton
bandgap. Analogous to the decay of a two-level atom in the photonic crystal,
we first point out that such an enhanced phenomenon allows one to examine
the non-Markovian dynamics of the QD exciton.   Besides, due to the one
dimensional propagating feature of nanowire surface plasmons, remote
entangled states can be generated via super-radiance and may be useful in
future quantum information processing.
\end{abstract}

\maketitle





The collective motions of an electron gas in a metal or semiconductor are
known as plasma oscillations. In the presence of surfaces, not only the bulk
modes are modified, but also surface modes can be created$^{1}$. When a
light wave strikes a metal surface, a surface plasmon polariton---a surface
electromagnetic wave that is coupled to plasma oscillations, can be excited.
Investigations of the dispersion relations of surface plasmons for different
geometries have been reported$^{2}$ since the 1970s. Recently, great
attention has been focused on the so-called plasmonics since surface
plasmons reveal strong analogies to light propagation in conventional
dielectric components$^{3}$. For examples, it is now possible to confine
them to subwavelength dimensions$^{4}$ and develop novel approaches for
waveguiding below the diffraction limit$^{5}$. The useful subwavelength
confinement, single mode operation$^{6}$, and relatively low power
propagation loss$^{7}$ of surface plasmon polaritons could be applied to
miniaturize photonic circuits$^{8}$. High surface plasmon field confinement
was also used to demonstrate an all-optical modulator$^{9}$.

Plasmon induced modification of the spontaneous emission (SE) rate is
naturally an extended issue. Arnoldus \emph{et al.} theoretically
investigated the atomic fluorescence and spontaneous decay near a metal
surface$^{10}$. Strong enhancement of fluorescence due to surface plasmons
was also observed$^{11}$. This enhanced fluorescence was considered as a
possible method to improve the quantum efficiency of light-emitting devices.
R. Paiella recently proposed tunable surface plasmons in silver-GaN multiple
layers to increase the radiative recombination rate and equivalently the
photoluminescence efficiency$^{12}$. Strong and coherent coupling between
individual optical emitters and guided plasmon excitations in conducting
nanowires at optical frequencies was also pointed out$^{13}$ and may be used
as a novel single-photon transistor$^{14}$.

In this work, we investigate the SE rate of a II-VI colloidal QD
(nanocrystals) exciton coupled to surface plasmons in a silver nanowire.
Radiative decay of a QD exciton into different modes of surface plasmons is
considered separately. The emission rate is found to be greatly enhanced at
certain values of QD exciton bandgap, which is similar to the band-edge
effect in photonic crystals. In addition, application of such a system in
generating remote entangled states is also pointed and may be useful in
future quantum information processing. 
\begin{figure}[th]
\includegraphics[width=8cm]{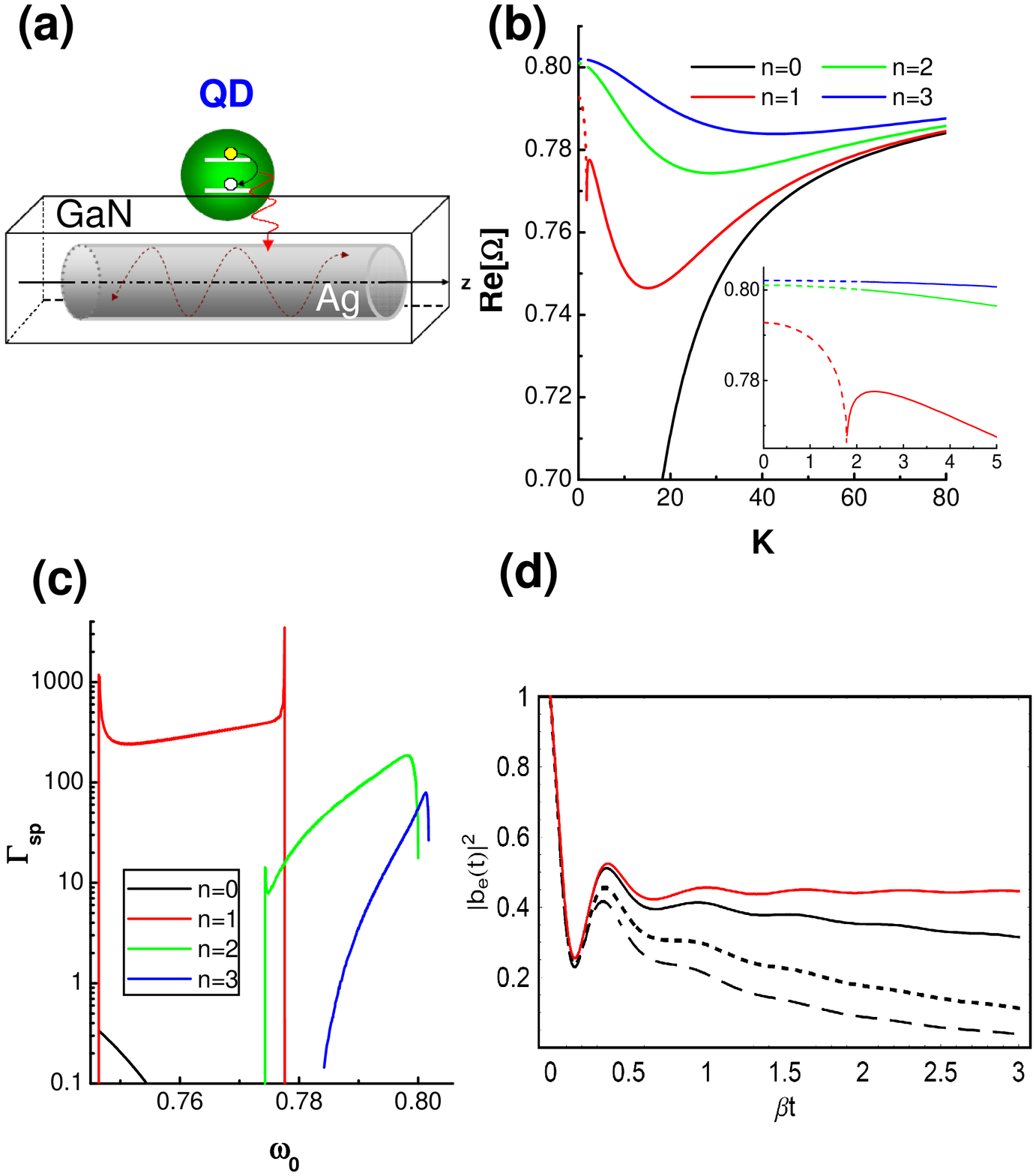}
\caption{{}(a) Schematic view of the model: a silver nano-wire is embedded
inside GaN material and a QD (quantum dot) is put on top of it. (b)
Dispersion relations of surface plasmons for the first few modes. The units
for vertical and horizontal lines are$\ \Omega =\protect\omega /\protect%
\omega _{p}$ and $K=k_{z}c/\protect\omega _{p}$ . (c) Corresponding
(Markovian) SE rates into surface plasmons. (d) Non-Markovian decay dynamics
of QD excitons for $\protect\delta =0.2\protect\beta $ (black line)$,$ $0.4%
\protect\beta $ (dotted line)$,$ and $0.8\protect\beta $ (dashed line). As $%
\protect\delta =0,$ the red line represents the result for the contribution
from $n=1$ mode.}
\end{figure}

\emph{Dispersion relations of surface plasmons.--}Consider now a colloidal
CdSe/ZnS QD near a cylindrical silver nanowire with radius $a$.\emph{\ }The
QD and nanowire are assumed to be separated by a GaN layer$^{15}$ as shown
in Fig. 1 (a). One of the main reasons to choose a CdSe/ZnS QD exciton as
the two-level emitter is that it is now possible to isolate single colloidal
QD and measure its exciton lifetime$^{16}$. The other reason is that its
exciton bandgap is around $2eV$ to $2.5eV$, depending on the size and
environment of the dot$^{17}$. The plasmon energy $\hbar \omega _{p}$ of
bulk silver is $3.76~eV$ with the corresponding saturation energy $\hbar
\omega _{p}/\sqrt{2}\approx 2.66eV$ in the dispersion relation$^{18}$. As we
shall see below, variations of the dispersion relations in energy just match
the exciton bandgap of colloidal CdSe/ZnS QDs.

Surface plasmon modes are created due to the nonzero local charge density on
the surface of a nanowire. The \emph{n-}th surface plasmon mode's components
of the electromagnetic field at the surface can be obtained by solving
Maxwell's equations in a cylindrical geometry ($\rho ~$and $\varphi $ denote
the radial and azimuthal coordinates, respectively) with the appropriate
boundary conditions$^{2}$: 
\begin{align}
& E_{\rho }=[\frac{ik_{z}}{K_{\xi }}\frac{d\psi _{n}^{\xi }(K_{\xi }\rho )}{%
d(K_{\xi }\rho )}A_{n}^{\xi }-\frac{\mu _{\xi }\omega n}{K_{\xi }^{2}\rho }%
\psi _{n}^{\xi }(K_{\xi }\rho )B_{n}^{\xi }]\phi _{n},  \notag \\
& E_{\varphi }=-[\frac{nk_{z}}{K_{\xi }^{2}\rho }\psi _{n}^{\xi }(K_{\xi
}\rho )A_{n}^{\xi }-\frac{i\mu _{\xi }\omega }{K_{\xi }}\frac{d\psi
_{n}^{\xi }(K_{\xi }\rho )}{d(K_{\xi }\rho )}B_{n}^{\xi }]\phi _{n},  \notag
\\
& E_{z}=[\psi _{n}^{\xi }(K_{\xi }\rho )A_{n}^{\xi }]\phi _{n},  \notag \\
& H_{\rho }=[\frac{n(K_{\xi }^{2}+k_{z}^{2})}{\mu _{\xi }\omega K_{\xi
}^{2}\rho }\psi _{n}^{\xi }(K_{j}\rho )A_{n}^{\xi }+\frac{ik_{z}}{K_{\xi }}%
\frac{d\psi _{n}^{\xi }(K_{\xi }\rho )}{d(K_{\xi }\rho )}B_{n}^{\xi }]\phi
_{n},  \notag \\
& H_{\varphi }=[\frac{i(K_{\xi }^{2}+k_{z}^{2})}{\mu _{\xi }\omega K_{\xi }}%
\frac{d\psi _{n}^{\xi }(K_{\xi }\rho )}{d(K_{\xi }\rho )}A_{n}^{\xi }-\frac{%
nk_{z}}{K_{\xi }^{2}\rho }\psi _{n}^{\xi }(K_{\xi }\rho )B_{n}^{\xi }]\phi
_{n},  \notag \\
& H_{z}=[\psi _{n}^{\xi }(K_{\xi }\rho )B_{n}^{\xi }]\phi _{n},
\end{align}%
with 
\begin{align}
& K_{\xi }^{2}=\omega ^{2}\epsilon _{\xi }(\omega )/c^{2}-k_{z}^{2}\text{ \ (%
}\xi =I\text{ or }O\text{)},  \notag \\
& \psi _{n}^{I}(K_{I}\rho )=J_{n}(K_{I}\rho ),~\psi
_{n}^{O}(K_{O}r)=H_{n}^{(1)}(K_{O}\rho ),  \notag \\
& \phi _{n}=exp(in\varphi +ik_{z}\emph{z}-i\omega t),  \notag
\end{align}%
where $J_{n}(K_{I}\rho )$ are $H_{n}^{(1)}(K_{O}\rho )$ are Bessel and
Hankel functions, respectively. \emph{I} (\emph{O}) stands for the component
inside (outside) the wire. The dielectric function is assumed as ${\epsilon }%
(\omega )=\varepsilon _{\infty }[1-\frac{\omega _{p}^{2}}{\omega (\omega
+i/\tau )}]$, where $\epsilon _{\infty }=9.6$ (for Ag), $\epsilon _{\infty
}=5.3$ (for GaN), and $\tau $ is the relaxation time due to ohmic metal loss$%
^{12}$. The magnetic permeabilities $\mu _{I,O}$ are unity everywhere since
we consider nonmagnetic materials here. $A_{n}^{\xi }$ and $B_{n}^{\xi }$
are constants to be determined by normalizing the electromagnetic field to
the vacuum fluctuation energy, $\int \epsilon (\left| E_{\rho }\right|
^{2}+\left| E_{\varphi }\right| ^{2}+\left| E_{z}\right| ^{2})d\mathbf{r}%
=\hbar \omega (\mathbf{k})$, and matching the boundary conditions. The
dispersion relations of the surface plasmons are thus obtained by solving
the following transcendental equation numerically:

\begin{align}
& S(k_{z},\omega )=  \notag \\
& [\frac{\mu _{I}}{K_{I}a}\frac{J_{n}^{\prime }(K_{I}a)}{J_{n}(K_{I}a)}-%
\frac{\mu _{O}}{K_{O}a}\frac{H_{n}^{(1)\prime }(K_{O}a)}{H_{n}^{(1)}(K_{O}a)}%
][\frac{(\omega /c)^{2}\varepsilon _{I}(\omega )}{\mu _{I}K_{I}a}\frac{%
J_{n}^{\prime }(K_{I}a)}{J_{n}(K_{I}a)}  \notag \\
& -\frac{(\omega /c)^{2}\varepsilon _{O}(\omega )}{\mu _{O}K_{O}a}\frac{%
H_{n}^{(1)\prime }(K_{O}a)}{H_{n}^{(1)}(K_{O}a)}]-n^{2}k_{z}^{2}[\frac{1}{%
(K_{O}a)^{2}}-\frac{1}{(K_{I}a)^{2}}]^{2}  \notag \\
& =0.
\end{align}%
\ 

The dispersion relations for various modes $n$ are shown in Fig. 1 (b) with
effective radii $R=0.1$. One unit of the effective radii $R$ ($\equiv \omega
_{p}a/c$) is roughly equal to $53.8~nm$. The behavior for the $n=0$ mode is
very similar to the two-dimensional case$^{18}$, i.e. $\Omega $ gradually
saturates with increasing wave vector $k_{z}$. This is because the fields
for the $n=0$ mode are independent of the azimuthal angle $\varphi $.
However, the behavior for the $n\neq 0$ modes are quite different. The first
interesting point are the discontinuities around $\omega /c\approx k_{z}$.
Further analysis shows that the solutions of $\omega $ are ``almost real''$%
^{19}$ as $k_{z}>Re[\omega ]/c$. Thus, the first Hankel function of order 
\emph{n}, $H_{n}^{(1)}(K_{\xi }\rho )$, decays exponentially. This means the
surface plasmons in this regime are confined on the surface (\textit{bound
modes}). For $k_{z}$ $<Re[\omega ]/c$, however, the solutions of $\omega $
are \emph{complex} as shown by the dashed lines in the inset of Fig. 1 (b). $%
H_{n}^{(1)}(K_{\xi }\rho )$ in this case is like a traveling wave with
finite lifetime (\textit{non-bound modes}).

Once the electromagnetic fields are determined, the spontaneous emission
(SE) rate, $\Gamma _{sp}$, of the QD excitons into bound surface plasmons
can be obtained via Fermi's golden rule. The SE rates of the first few modes
($n=0,1,2,3$) are shown in Fig. 1 (c) with effective radii $R=0.1$. In
plotting the figures, the distance between the dot and the wire surface is
fixed as $d=10.76~nm$. The novel feature is that the SE rate approaches
infinity at certain values of the exciton bandgap $\omega _{0}$.
Mathematically, one might think that at these values the corresponding
slopes of the dispersion relation are zero. Physically, however, this
infinite rate is not reasonable since it's based on perturbation theory.
Therefore, one has to treat the dynamics of the exciton around these values
more carefully, i.e. the \emph{Markovian} SE rate is not enough. One has to
consider the \emph{non-Markovian} behavior around the band-edge, which means
the band abruptly appears/disappears across certain values of $\omega $.

\emph{Non-Markovian dynamics of QD excitons.--}To obtain the non-Markovian
dynamics of the exciton, we first write down the Hamiltonian of the system
in the interaction picture (with the rotating wave approximation),

\begin{eqnarray}
H_{ex-sp} &=&\sum_{n,k_{z}}\hbar \Delta _{n,k_{z}}\widehat{a}%
_{n,k_{z}}^{\dag }\widehat{a}_{n,k_{z}}  \notag \\
&&+i\hbar \sum_{n,k_{z}}(g_{n,k_{z}}\widehat{a}_{n,k_{z}}^{\dag }\sigma
_{\downarrow \uparrow }-g_{n,k_{z}}^{\ast }\widehat{a}_{n,k_{z}}\sigma
_{\uparrow \downarrow }),
\end{eqnarray}%
where $\sigma _{ij}=\left| i\right\rangle \left\langle j\right| $($%
i,j=\uparrow ,\downarrow $) are the atomic operators; $\widehat{a}_{n,k_{z}}$
and $\widehat{a}_{n,k_{z}}^{\dag }$ are the radiation field (surface
plasmon) annihilation and creation operators; $\Delta _{n,k_{z}}=\omega
_{n,k_{z}}-\omega _{0}$ is the detuning of the radiation mode frequency $%
\omega _{n,k_{z}}$ from the excitonic resonant \ frequency $\omega _{0}$,
and $g_{n,k_{z}}=\overrightarrow{\mu }\cdot \overrightarrow{E}_{n,k_{z}}$ is
the atomic field coupling. Here, $\overrightarrow{\mu }$ and $%
\overrightarrow{E}_{n,k_{z}}$ denote the transition dipole moment of the
exciton and the electric field, respectively.

Assuming there is an exciton in the dot with no plasmon excitation in the
wire initially, the wavefunction of the system then has the form%
\begin{equation}
\left| \psi (t)\right\rangle =b_{e}(t)\left| \uparrow ,0\right\rangle
+\sum_{n,k_{z}}b_{n,k_{z}}(t)\left| \downarrow ,1_{n,k_{z}}\right\rangle
e^{-i\Delta _{n,k_{z}}t}.
\end{equation}%
The state vector $\left| \uparrow ,0\right\rangle $ describes an exciton in
the dot and no plasmons present, whereas $\left| \downarrow
,1_{n,k_{z}}\right\rangle $ describes the exciton recombination and a
surface plasmon emitted into mode $k_{z}$. With the time-dependent Schr\"{o}%
dinger equation, the solution of the coefficient $b_{e}(t)$ in $z$-space is
straightforwardly given by

\begin{equation}
\widetilde{b}_{e}(z)=[z+\sum_{n,k_{z}}g_{n,k_{z}}g_{n,k_{z}}^{\ast }\frac{1}{%
z+i(\omega -\omega _{0})}]^{-1}.
\end{equation}%
In principle, $b_{e}(t)$ can be obtained by performing a numerical inverse
Laplace Transformation to Eq. (5).

To grasp the main physics and without loss of generality, we focus on the
values of $\omega _{0}$ close to one particular local extremum. In this
case, the dispersion relation for this particular $n$ mode around the
extremum can be approximated as $\omega _{k_{z}}=\omega _{n,c}\pm
A_{n}(k_{z}-k_{n,c})^{2},$where the extremum is located at ($k_{n,c},\omega
_{n,c}$). The $+/-$ sign represents the approximate curve for the local
minimum/maximum of the dispersion relation. Once we make such an
approximation, the radiative dynamics of the QD exciton is just like that of
a two-level atom in a photonic crystal$^{20}$with 
\begin{equation}
\widetilde{b}_{e}(z)\approx \left\{ 
\begin{array}{cc}
\frac{\left| \overrightarrow{\mu }\cdot \overrightarrow{E}%
_{n,k_{z}=k_{n,c}}\right| ^{2}}{z-\gamma /2-\frac{(-1)^{3/4}\pi }{\sqrt{A_{n}%
}\sqrt{z-i\delta }}},\text{for local minimums} &  \\ 
\frac{\left| \overrightarrow{\mu }\cdot \overrightarrow{E}%
_{n,k_{z}=k_{n,c}}\right| ^{2}}{z-\gamma /2+\frac{(-1)^{1/4}\pi }{\sqrt{A_{n}%
}\sqrt{z-i\delta }}},\text{for local maximums} & 
\end{array}%
,\right.
\end{equation}%
where $\delta =$ $\omega _{0}-\omega _{n,c}$ is the detuning and $\gamma $
is the decay rate contributed from other modes.

The coefficient $b_{e}(t)$ can now be obtained by performing the Laplace
transformation to Eq. (5)$^{20}$. The black, dotted, and dashed lines in
Fig. 1 (d) represent the decay dynamics of the QD excitons for different
detunings: $\delta =0.2\beta ,$ $0.4\beta ,$ $0.8\beta ,$ respectively.
Here, $\beta $ is the decay rate of the QD exciton in free space. In
plotting the figure, $\omega _{0}$ is chosen to be close to the local
minimum of the dispersion relation of the $n=1$ mode. The radius of the wire
and the wire-dot separation are identical to those in Fig. 1 (b). As can be
seen, there exists oscillatory behavior in the decay profile, demonstrating
that decay dynamics around the local extremums is non-Markovian. If one
considers only the contribution from the $n=1$ mode and set the detuning $%
\delta =0$, the probability amplitude would saturate to a steady limit as
show by the red line. This \emph{quasi-dressed} state is an analogous of
Rabi-oscillation in cavity quantum electrodynamics, and also appears in the
systems of photonic crystals$^{20}$. 
\begin{figure}[th]
\includegraphics[width=8cm]{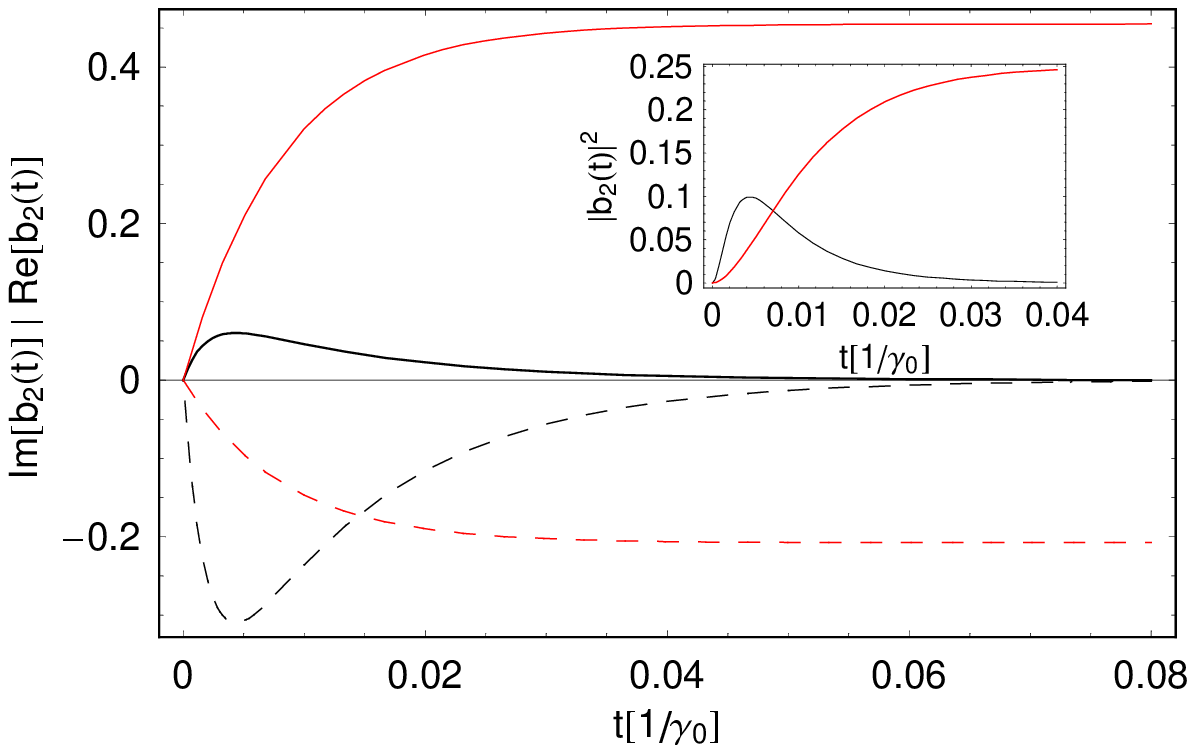}
\caption{{}Variations of Re[$b_{2}(t)$] (solid lines), Im[$b_{2}(t)$]
(dashed lines), and $\left| b_{2}(t)\right| ^{2}$ (inset) as functions of
time for $\protect\omega _{0}=0.602\protect\omega _{p}$ (red lines) and $%
0.748\protect\omega _{p}$ (black lines). In plotting the figure, the
inter-dot distance $z_{0}$ is set equal to $0.35(\protect\omega _{p}a/c)$
with radius $R=0.1$.}
\end{figure}

\emph{Application in entanglement generation.--}Let us now put another QD
close to the wire, the interaction between the wire and QDs can now be
written as 
\begin{equation}
H^{\prime }=\sum_{n,k_{z}}g_{n,k_{z}}(\sigma _{1+}+\sigma
_{2+}e^{ik_{z}z_{0}})\widehat{a}_{n,k_{z}}+h.c.,
\end{equation}%
where $a_{n,k_{z}}$ is the surface-plasmon operator, $\sigma _{j+}$ is the
creation operator of the $j$-th QD. Note that in Eq. (7) we have assumed no
detuning and the two dots have the same separation from the metal wire.
Since the propagating modes are along the $z$-direction only, the phase
difference acquired by the second dot is $ik_{z}z_{0}$, where $z_{0}$ is the
separation between the two dots. If one further assumes that only QD-1 is
initially excited, the state vector \ of the system can be written as

\begin{equation}
\left| \Psi (t)\right\rangle =b_{1}(t)\left| \uparrow \downarrow
;0\right\rangle +b_{2}(t)\left| \downarrow \uparrow ;0\right\rangle
+\sum_{n,k_{z}}b_{n,k_{z}}(t)\left| \downarrow \downarrow
;1_{n,k_{z}}\right\rangle
\end{equation}%
with $b_{1}(0)=1$ and $b_{2}(0)=b_{n,k_{z}}(0)=0$. Here, $\left| \uparrow
\downarrow ;0\right\rangle $ ($\left| \downarrow \uparrow ;0\right\rangle $)
means that QD-1(-2) is excited, while $\left| \downarrow \downarrow
;1_{n,k_{z}}\right\rangle $ represents that both the QDs are deexcited with
the presence of single surface-plasmon. If we let the exciton band-gap $%
\omega _{0}$ far away from the band-edge, $b_{1}(t)$ and $b_{2}(t)$ can be
obtained easily by solving the time-dependent Schr\"{o}dinger equation.

Fig. 2 shows the time variations of Re[$b_{2}(t)$], Im[$b_{2}(t)$], and $%
\left| b_{2}(t)\right| ^{2}$ for different values of the exciton bandgap.
For $\omega _{0}=0.748\omega _{p}$, the population of the second dot
vanishes quickly as seen from the black lines. For $\omega _{0}=0.602\omega
_{p}$, however, it approaches (quasi-)stationary limit$^{21}$ as shown by
the red lines. This is because, for the later case, only $n=0$ mode
contributes to the decay rate $\Gamma _{sp}$. In this case, the populations
of the dots can be analytically written as

\begin{equation}
\left\{ 
\begin{array}{cc}
b_{1}(t)=e^{-2\Gamma _{sp}t}(1+e^{2\Gamma _{sp}t})/2, &  \\ 
b_{2}(t)=e^{-ik_{0}z_{0}-2\Gamma _{sp}t}(-1+e^{2\Gamma _{sp}t})/2. & 
\end{array}%
\right.
\end{equation}%
From Eq. (9), one realizes that there is always 50\% chance for the two dots
to evolve into the state: $\left| \uparrow \downarrow \right\rangle +$ $%
e^{-ik_{0}z_{0}}\left| \downarrow \uparrow \right\rangle $. It means that,
for example, the singlet [triplet] entangled state can be created if $%
k_{0}z=(2m+1)\pi $ [$2m\pi $] with $m$ being an integer. The advantage is
that the entangled states can be generated with a remote sense, such that
one can manipulate/control one qubit without affecting another. One should
be reminded again such an entanglement is generated from the collective
decay (super-radiance) in one dimension$^{22}$, not from the trivial
single-mode Rabi coupling, since the decay rate $\Gamma _{sp}$ is still
present in Eq. (9).

In summary, we have shown that radiative decay of colloidal QD excitons into
surface plasmons can be greatly enhanced at certain values of the exciton
bandgap. The enhancement is due to zero-slope in dispersion relation, and
one has to treat the decay dynamics with a non-Markovian way. In addition,
an idea of creating remote entangled states between two QDs is also proposed
and can be tested with current technology$^{23}$.

\subsection{Acknowledgments}

We would like to thank Prof. T. Brandes at Technische Universit\"{a}t Berlin
and Dr. J. Taylor at MIT for helpful discussions. This work is supported
partially by the National Science Council, Taiwan under the grant numbers
NSC 96-2112-M-009-021 and 95-2112-M-006-031-MY3.


\begin{thebibliography}{99}
\bibitem{science} From the numerical results, the imaginary parts of $\omega 
$ for the bound modes are actually very small ($10^{-4}\symbol{126}$ $%
10^{-5} $ of the real parts).
\bibitem{21} In fact, the exciton eventually decays into free space. Since
the coupling to the surface plasmons is so strong, one can roughly neglecte
the effect from vacuum fluctuations in the regime of $t<1/\gamma _{0}$.
\end{thebibliography}
\end{document}